\documentclass[prc,aps,preprint,showkeys,lengthcheck,twocolumn,nofootinbib,notitlepage,floatfix,superscriptaddress, nolongbibliography]{revtex4-2}
\usepackage{graphics,graphicx}
\usepackage{amsmath, amssymb}
\usepackage{multirow}
\usepackage{color}
\usepackage{verbatim}
\usepackage{hyperref}
\usepackage[normalem]{ulem}  % \sout{old text} for strikeout
\usepackage{ulem}
\usepackage{color}
\usepackage{cancel}
\usepackage{mathtools}
\usepackage{epstopdf}
\usepackage{soul}
\renewcommand\sout{\bgroup\color{blue} \ULdepth=-.5ex \ULset}
\def\slashchar#1{\setbox0=\hbox{$#1$}  % set a box for #1
\dimen0=\wd0     % and get its size
\setbox1=\hbox{/} \dimen1=\wd1  % get size of /
\ifdim\dimen0>\dimen1   % #1 is bigger
\rlap{\hbox to \dimen0{\hfil/\hfil}} % so center / in box
#1     % and print #1
\else     % / is bigger
\rlap{\hbox to \dimen1{\hfil$#1$\hfil}} % so center #1
/      % and print /
\fi}

\newcommand{\dd}{\mathrm{d}}

\newcommand{\eps}{\epsilon}

\begin{document}

\title{Magnetic effects  in the Hadron Resonance Gas}

\date{\today}
\author{Micha\l{} Marczenko}
\email{michal.marczenko@uwr.edu.pl}
\affiliation{Incubator of Scientific Excellence - Centre for Simulations of Superdense Fluids, University of Wroc\l{}aw, plac Maksa Borna 9, PL-50204 Wroc\l{}aw, Poland}
\author{Micha\l{} Szyma\'nski}
\affiliation{Institute of Theoretical Physics, University of Wroc\l{}aw, plac Maksa Borna 9, PL-50204 Wroc\l{}aw, Poland}
\author{Pok Man Lo}
\affiliation{Institute of Theoretical Physics, University of Wroc\l{}aw, plac Maksa Borna 9, PL-50204 Wroc\l{}aw, Poland}
\author{Bithika Karmakar}
\affiliation{Incubator of Scientific Excellence - Centre for Simulations of Superdense Fluids, University of Wroc\l{}aw, plac Maksa Borna 9, PL-50204 Wroc\l{}aw, Poland}
\author{Pasi Huovinen}
\affiliation{Incubator of Scientific Excellence - Centre for Simulations of Superdense Fluids, University of Wroc\l{}aw, plac Maksa Borna 9, PL-50204 Wroc\l{}aw, Poland}
\author{Chihiro Sasaki}
\affiliation{Institute of Theoretical Physics, University of Wroc\l{}aw, plac Maksa Borna 9, PL-50204 Wroc\l{}aw, Poland}
\affiliation{International Institute for Sustainability with Knotted Chiral Meta Matter (WPI-SKCM$^2$), Hiroshima University, Higashi-Hiroshima, Hiroshima 739-8526, Japan}
\author{Krzysztof Redlich}
\affiliation{Institute of Theoretical Physics, University of Wroc\l{}aw, plac Maksa Borna 9, PL-50204 Wroc\l{}aw, Poland}
\affiliation{Polish Academy of Sciences PAN, Podwale 75, 
PL-50449 Wroc\l{}aw, Poland}

\begin{abstract}
We discuss the modeling of the hadronic phase of QCD at finite magnetic field in the framework of hadron resonance gas (HRG). We focus on the statistical description of particle yields that include contribution from resonance decays. We demonstrate that the swift increase in the number of protons with magnetic field predicted in the HRG is due to the assumption of structureless resonances. We discuss fluctuations of conserved charges and show that at present the qualitative comparison of the model predictions with the Lattice QCD data should be treated with care. We also discuss the principle of detailed balance which allows us to study the magnetic field dependence of neutral resonances.

\end{abstract}
\maketitle

\section{Introduction}

Magnetic fields generated in non-central heavy-ion collisions (HICs) are among the strongest in the universe~\cite{Kharzeev:2007jp, Skokov:2009qp, Bzdak:2011yy}. Since quarks and antiquarks are electrically charged and influenced by magnetic field, understanding the interplay between the magnetic field and the strong interactions governed by quantum chromodynamics (QCD) is crucial for properly comprehending non-central HICs. Understanding the impact of an extremely strong magnetic field on the strongly interacting matter is also relevant for studying magnetars~\cite{Duncan:1992hi} and the early universe~\cite{Vachaspati:1991nm}. The magnetic field induces a variety of interesting phenomena in the QCD matter, such as anomaly-induced transport phenomena, e.g., Chiral Magnetic Effect (CME)~\cite{Fukushima:2008xe} and Chiral Vortical Effect (CVE)~\cite{Landsteiner:2012kd}, charged $\rho$ meson condensation~\cite{Chernodub:2010qx} and modifications of QCD Debye mass~\cite{Bandyopadhyay:2017cle}, QCD equation of state (EoS)~\cite{Bali:2014kia, Bandyopadhyay:2017cle, Brandt:2024blb, MarquesValois:2023ehu} and dilepton production rate~\cite{Sadooghi:2016jyf, Das:2021fma}.

Investigation of QCD in the background of magnetic fields is also interesting from the theoretical standpoint since it can be simulated from the first principle lattice QCD (LQCD) methods even for large fields in contrast to finite baryon chemical potential where the simulations are marred by the numerical sign problem. LQCD simulations indicate a nontrivial interplay between the magnetic field and strong interactions, such as magnetic catalysis at low temperatures~\cite{DElia:2011koc, Bali:2012zg}, and inverse magnetic catalysis~\cite{Bruckmann:2013oba, DElia:2018xwo, Endrodi:2019zrl} which influences the QCD phase diagram. These effects have also been widely studied using several other formalisms, {\it{viz.}} Nambu-Jona-Lasinio (NJL) model, Polyakov improved NJL (PNJL) model~\cite{Gatto:2012sp}, quark-meson (QM) model~\cite{Gatto:2012sp}, and gauge-gravity duality~\cite{Bergman:2012na}. However, the inverse magnetic catalysis is not captured in commonly used chiral models~\cite{Andersen:2014xxa}. Nevertheless, it can be generated by considering additional in-medium effects (see e.g.~\cite{Lo:2021buz}). It has also been found that the pseudocritical temperature of the QCD crossover decreases with increasing magnetic field (see, e.g., Refs.~\cite{Bali:2011qj, Endrodi:2015oba, DElia:2018xwo}), with the possible existence of a critical point at large field strength~\cite{Endrodi:2015oba, DElia:2021yvk}. For recent reviews of QCD in external magnetic field, see~\cite{Andersen:2014xxa, Miransky:2015ava, Fukushima:2018grm, Hattori:2023egw}. Recently, the fluctuations of conserved charges in the magnetic field background have attracted considerable interest and they have been evaluated using LQCD methods~\cite{Ding:2021cwv, Ding:2023bft, MarquesValois:2023ehu, Astrakhantsev:2024mat}. It has been observed that a finite magnetic field has a nontrivial effect on these fluctuations. They are of direct phenomenological interest because they are sensitive to in-medium degrees of freedom and can be measured experimentally. 

The hadron resonance gas (HRG) model has been successful in describing the LQCD data on the EoS and fluctuations of conserved charges at vanishing magnetic field~\cite{HotQCD:2014kol, Bollweg:2021vqf} as well as the hadron yields from heavy-ion collisions~\cite{Andronic:2017pug}. In~\cite{Endrodi:2013cs}, the HRG model has been generalized to include the effects of finite magnetic field. The model has been used to study the impact of magnetic field on the QCD EoS~\cite{Endrodi:2013cs, Bali:2014kia}, fluctuations of conserved charges~\cite{Bhattacharyya:2015pra, Mohapatra:2017zrj, Ding:2023bft, Ding:2022uwj} as well as other effects, such as transport coefficients~\cite{Das:2019pqd, Dash:2020vxk}. The behavior of spin-$0$ and spin-$1/2$ particles in an external magnetic field is well understood and involves the Landau quantization in the directions perpendicular to the magnetic field~\cite{Landau:1977}. However, when the same prescription is applied to higher-spin resonances, it leads to imaginary dispersion relations at strong fields as well as instabilities in pressure~\cite{Endrodi:2013cs}. This is due to the bold assumption of structureless resonance states in the HRG model. This suggests that the treatment of higher-spin resonances in the HRG model is doubtful and requires further investigation.

In this work, we study the fluctuations of conserved charges and particle yields at a finite magnetic field using the HRG model. We find that the magnetic field enhances the baryon yields, while $\pi^{\pm}$ and $K^{\pm}$ yields are weakly affected. With our choice of resonances included in HRG, we find that the model underestimates the LQCD baryon number fluctuations at a weak magnetic field while overestimating them at large $eB$, the latter originating from the divergences induced by higher-spin resonances. We also gauge the impact of the magnetic field on neutral resonances by implementing detailed balance conditions. We compare the proton and pion yields obtained using HRG and our detailed balance requirement and observe a substantial difference, indicating the need for further research on the impact of strong magnetic field on resonance properties.  

This paper is organized as follows: In Sec.~\ref{sec:hrg} we introduce the HRG thermodynamics at finite temperature and magnetic field and show the numerical results. We discuss the effect of magnetic field on particle densities and yields, and fluctuations of conserved charges in conjunction with the LQCD data. In Sec.~\ref{sec:det_balance}, we show how the external magnetic field affects neutral particle densities considering detailed balance conditions. 
Finally, Sec. \ref{sec:conclusions} is devoted to summary and conclusions.

\section{Hadronic Matter in finite magnetic field}
\label{sec:hrg}

In a phenomenological description of hadronic matter, one needs to identify the relevant degrees of freedom and their interactions. In the confined phase of QCD, the medium is composed of hadrons and their resonances. In its simplest version, the hadron resonance gas (HRG) model assumes that the constituents of the medium are independent and point-like~\cite{Braun-Munzinger:2003pwq}. This effectively neglects their widths and interactions. Consequently, the pressure in the HRG model is approximated by the sum over partial pressures of hadrons and their resonances, treated as noninteracting particles,
\begin{equation}\label{eq:hrg_pressure}
    P=\sum\limits_{i} P_i \rm ,
\end{equation}
where $i$ goes through all strange and non-strange hadrons and their resonances listed in the Particle Data Group summary tables~\cite{ParticleDataGroup:2022pth}\footnote{In this work, we include established mesons and baryons with three- and four-star rating.}. We note that the thermodynamic pressure $P$ contains all the relevant information about the medium through the mass and quantum numbers of hadrons. Thus, it allows for the study of different thermodynamic observables, including particle numbers and fluctuations of conserved charges.

The partial pressures in Eq.~\eqref{eq:hrg_pressure} are given as
\begin{equation}\label{eq:p_i}
    P_{i} = \pm \gamma_i T \int \frac{\dd^3p}{(2\pi)^3}\; \log\left(1\pm f_i\right)\textrm,
\end{equation}
where $\gamma_i$ is the spin degeneracy factor and 
\begin{equation}\label{eq:distr}
    f_i = \frac{1}{e^{\beta (\eps_i-\mu_i)} \pm 1}
\end{equation}
is the distribution function, $\eps_i=\sqrt{\Vec{p}^2+m_i^2}$ is the dispersion relation and \mbox{$\mu_i=B_i \mu_B+Q_i\mu_Q+S_i\mu_S$} is the chemical potential of the particle. The upper (lower) signs in Eqs.~\eqref{eq:p_i} and \eqref{eq:distr} refer to fermions (bosons). 

\begin{figure}[t!]
    \centering
    \includegraphics[width=\columnwidth]{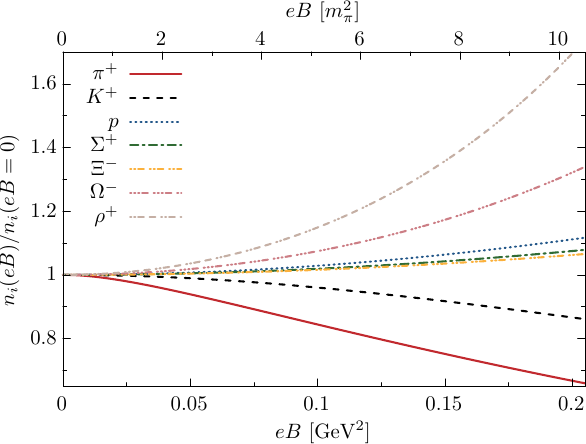}
    \caption{Particle densities normalized to densities at vanishing magnetic field at $T=0.155~$GeV as functions of magnetic field $eB$.}
    \label{fig:density_eB}
\end{figure}

In the presence of a constant external magnetic field $eB$ pointing along the $z$ direction, the system undergoes Landau quantization in the $xy$ plane~\cite{Andersen:2014xxa, Miransky:2015ava}. Consequently, the dispersion relation for a charged particle $(Q>0)$ becomes
\begin{equation}
\label{eq:energy_B}
    \eps = \sqrt{p_z^2 + m^2 + 2|Q|B\left(l + \frac{1}{2} - s_z\right)}\textrm,
\end{equation}
where $s_z$ is the $z$-component of the particle's spin and $l \in \{0,1,2,\ldots \}$ numbers the Landau levels. We note that the dispersion relation in Eq.~\eqref{eq:energy_B} is exact for structureless spin-$0$ and spin-$1/2$ particles~\cite{Andersen:2014xxa}. On the other hand, it has been found that spin-$3/2$ particles, governed by the Rarita-Schwinger equation, exhibit a non-causal behavior in the presence of external magnetic field~\cite{Velo:1969bt, Johnson:1960vt, dePaoli:2012eq}, which can be fixed by adding a non-minimal interaction term to the Lagrangian~\cite{Ferrara:1992yc, Porrati:2009bs}. Such a procedure leads to a dispersion relation similar to Eq.~\eqref{eq:energy_B}.

We stress that Eq.~\eqref{eq:energy_B} neglects the compositeness of resonances, that is, the possibility of decaying to daughter particles. This assumption becomes questionable when the scale of magnetic field resolves the structure of hadron states, i.e., $\sqrt{|Q|B} > m_\pi$. Problems become more apparent for higher spin states, where the effective mass of particles can be substantially reduced, which eventually leads to a complex dispersion relation (signaling instability). Resolving these issues would require a more sophisticated treatment of the decay dynamics~\cite{Bandyopadhyay:2016cpf, Ghosh:2017rjo}. 

Furthermore, in general, the masses of hadrons should be affected by the presence of an external magnetic field. This has been studied in the LQCD~\cite{Hidaka:2012mz, Bali:2017ian, Endrodi:2019whh, Ding:2022tqn} and effective models~\cite{Coppola:2019uyr, Avancini:2016fgq, Ghosh:2016evc, Ayala:2020muk, Coppola:2023mmq}. For example, the continuum-extrapolated mass of $\pi^0$  from the first-principle LQCD calculations decreases with increasing magnetic field. However, in the range of magnetic fields considered in our work~\cite{Bali:2017ian}, the change is less than 10$\%$. The magnetic-field-dependent masses of nucleons and $\Sigma$ baryons at zero temperature were also calculated in LQCD~\cite{Endrodi:2019whh}. Again,  in the range of magnetic fields considered in our work, they are changing only mildly. Beyond this calculation, it is challenging to estimate the effect of the magnetic field on the masses of other particles, in particular hadronic resonances. The systematic implementation of the mass shifts for individual hadronic resonances requires a more elaborate treatment of interactions~\cite{Schenk:1991xe,Jeon:1998zj,Lo:2022nhk} going beyond the current hadron resonance gas model. Reliable results are limited to only a few channels and the rest remain highly model-dependent. Therefore, we do not account for these effects in this work.

When the energies of particles are given by Eq.~\eqref{eq:energy_B}, the phase-space integral of a charged particle acquires the following form~\cite{Chakrabarty:1996te, Fraga:2008qn}
\begin{equation}
\label{eq:sum_B}
    \gamma\int \frac{\dd^3p}{(2\pi)^3} \rightarrow \frac{|Q|B}{2\pi^2}\sum_{s_z}\sum_{l=0}^\infty  \int\limits_0^\infty \dd p_z \textrm,
\end{equation}
where $\gamma$ is the spin degeneracy factor and $Q$ is the electric charge of the particle.
The presence of a homogeneous background magnetic field breaks the spherical symmetry, causing the pressure transverse to the magnetic field ($P^\perp$) to be different than the pressure parallel to it ($P^\parallel$). As demonstrated in Ref.~\cite{Strickland:2012vu} for a Fermi gas, the system satisfies the canonical relations $\Omega = - P^\parallel$, and $P^\perp = P^\parallel - BM$, where $M = -\partial \Omega / \partial B = \partial P^\parallel / \partial B$ is the magnetization. $P^\parallel$ can be obtained by applying prescription \eqref{eq:sum_B} to Eq.~\eqref{eq:p_i}, 
\begin{equation}\label{eq:landau}
    P^\parallel_{i} =\pm T \frac{|Q_i|B}{2\pi^2}\sum_{s_z}\sum_{l=0}^\infty  \int\limits_0^\infty \dd p_z\; \log\left(1\pm f_i\right)\,.
\end{equation}
However, the contribution of the magnetization current density to the transverse pressure~\cite{Blandford} is not considered in this formalism. If this contribution is taken into account, the pressure becomes isotropic~\cite{Blandford} (See also Refs.~\cite{Potekhin:2011eb, Ferrer:2011ze}). Nevertheless, it has been shown~\cite{Bali:2013esa} that whether one should include this contribution or not, depends on the formal definition of pressure as a derivative of the free energy $F$, i.e., whether the magnetic field or the flux of the field is kept constant while taking the derivative. In both cases observables such as the energy density, entropy, number density, and fluctuations of conserved charges are equal and do not depend on the scheme used to compute the pressure~\cite{Endrodi:2024cqn}. Therefore, we use Eq.~\eqref{eq:landau} to define pressure of the hadron gas and omit the $\parallel$ symbol in the remainder of this work.

\begin{figure}[t!]
    \centering
    \includegraphics[width=\columnwidth]{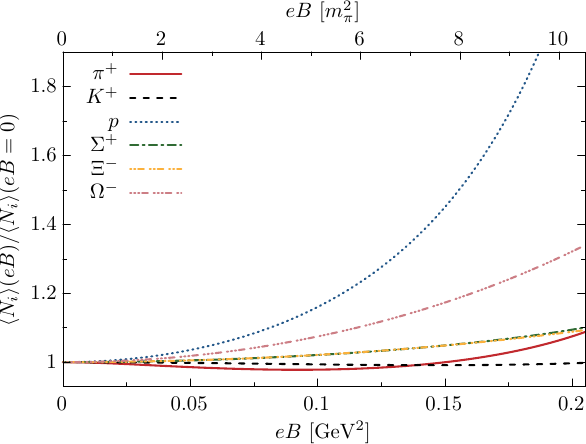}
    \caption{Particle yields normalized to the yields at vanishing magnetic field at $T=0.155~$GeV as functions of the magnetic field $eB$.}
    \label{fig:density_decay_eB}
\end{figure}

The pressure of a finite magnetic field contributes to the other thermodynamic quantities as well, and the vacuum part of the thermodynamic pressure must be properly renormalized~\cite{Endrodi:2013cs}. In this work, we focus on the densities and fluctuations of conserved charges which do not depend on the vacuum part of the pressure. Thus, we omit it in our considerations. We also note that the magnetic field does not distinguish between positive and negative electric charges (cf. Eqs.~\eqref{eq:energy_B} and~\eqref{eq:landau}).

The particle density is defined as 
\begin{equation}\label{eq:density}
    n_i = \frac{\langle N_i \rangle}{V} = \frac{\partial P_i}{\partial \mu_i}\Bigg|_{T} \rm,
\end{equation}
where $\langle N_i\rangle$ is an average number of particle $i$ in volume $V$. To quantify the change of the medium composition due to the presence of a finite magnetic field, we study the densities (normalized to their densities at vanishing magnetic field) of selected hadrons at fixed temperature $T=0.155~\rm GeV$. This is depicted in Fig.~\ref{fig:density_eB}. The densities of $p$, $\Sigma^+$, and $\Xi^-$ generally increase with the magnetic field. This is attributed to the lowest Landau level being magnetic-field independent and hence its contribution grows linearly with $eB$ (cf. Eqs.~\eqref{eq:energy_B} and~\eqref{eq:landau}). In contrast to this, the densities of $\pi^-$ and $K^-$ are exponentially suppressed because in this case the energies of all Landau levels increase with magnetic field. We note that there is a clear mass ordering for spin-$1/2$ baryons and spin-$0$ mesons: the lighter the particle, the larger the effect. The exceptions to this mass ordering are the higher-spin particles, $\rho^+$ and $\Omega^-$, whose increase is much greater than for other particle species. This swift increase is caused by the contribution of the lowest Landau level being negative for these particles, see Eq.\eqref{eq:energy_B}. 
As mentioned, the extrapolation of Eq.~\eqref{eq:energy_B} to higher-spin states requires the assumption of structureless hadrons, which can be questionable depending on the strength of magnetic field and the structure of the resonance in question.

\begin{figure}[t!]
    \centering
    \includegraphics[width=\columnwidth]{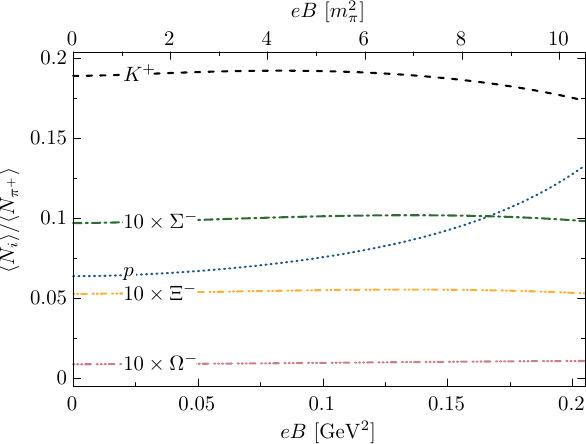}
    \caption{Particle yields normalized to the $\pi^+$ yield at $T=0.155~$GeV as functions of magnetic field $eB$.}
    \label{fig:density_decay_pion_eB}
\end{figure}

The HRG model can be used to determine the hadron yields, assuming thermal and chemical equilibrium between all stable hadrons and resonances (see, e.g.,~\cite{Andronic:2017pug, Andronic:2005yp, Andronic:2020iyg}). This is done by accounting for the resonance decays into lighter particles. The average number $\langle N_i\rangle$ of particle $i$ in volume $V$ is given as~\cite{Braun-Munzinger:2003pwq}
\begin{equation}\label{eq:yield}
    \langle N_i\rangle = \langle N^{\rm th}_i\rangle + \sum_{r \rightarrow i} \Gamma_{r\rightarrow i} \langle N_r\rangle \rm,
\end{equation}
where $\langle N^{\rm th}_i \rangle = Vn_i$ is the average thermal number defined through Eq.~\eqref{eq:density}, the sum in the second term is over all decays with particle $i$ in the final state, and $\Gamma_{r \rightarrow i}$ is the branching ratio of the resonance $r$ to particle $i$.

In Fig.~\ref{fig:density_decay_eB}, we show the particle yields normalized to the yields at the vanishing magnetic field. At small values of $eB$, the $\pi^+$ and $K^+$ yields decrease with increasing $eB$, but at stronger magnetic field, their yields increase with increasing $eB$. Nevertheless, at $eB \lesssim 0.2~\rm GeV^2$, $\pi^+$ and $K^+$ yields change only a few percent.  However, at the same values of $eB$, proton yields increase substantially, due to decaying $\Delta$ resonances. We warn the reader again about the uncertainties related in the description of high spin ($s \geq 1$) states and note that the weak dependence of $\Sigma$ and $\Xi$ yields on the magnetic field may be a result of the way fewer known strange than non-strange resonances. 

In the literature, the observed particle yields are often discussed in terms of particle ratios instead of actual yields to avoid dependency on system size. To facilitate the discussion in terms of particle ratios, we show the particle yields normalized to the yield of $\pi^+$ in Fig.~\ref{fig:density_decay_pion_eB}. Again, the prominent feature is the behavior of protons: the $p/\pi$ ratio increases substantially with the increasing strength of the magnetic field. We note that even if the $\Omega^+$ yield increases way faster than the pion yield in Fig.~\ref{fig:density_decay_eB}, it is still so tiny compared to the pion yield that the change in the ratio cannot be seen at the scale of Fig.~\ref{fig:density_decay_pion_eB}.

\begin{figure}[t!]
    \centering
    \includegraphics[width=\columnwidth]{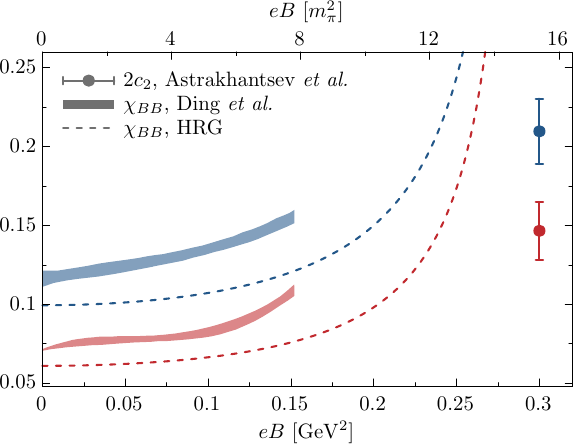}
    \caption{Net-baryon number susceptibility for $T=0.145$ (red) and $0.155~$GeV (blue) as a function of magnetic field $eB$.}
    \label{fig:c2_comp}
\end{figure}

Now we turn to the fluctuations of conserved charges. In a thermal medium, the second-order fluctuations and correlations of conserved charges are quantified by the generalized susceptibilities,
\begin{equation}
    \chi_{ij}=\frac{\partial^2 (P/T^4)}{\partial(\mu_i/T)\partial(\mu_j/T)}\Bigg|_{T}.
\end{equation}
where $i,j\in \lbrace B,~Q,~S\rbrace$. The susceptibilities are observables sensitive to the quantum numbers of medium constituents. Thus, they can be used to identify the contributions of different species of particles to QCD thermodynamics~\cite{Ejiri:2005mk, Ejiri:2005wq}.

For a sufficiently small net-baryon density, the EoS can be expanded in the Taylor series. Under the assumption of strangeness neutrality, the pressure is expanded as~\cite{Astrakhantsev:2024mat}
\begin{equation}
    \frac{P}{T^4} \approx c_0 + c_2 \left(\frac{\mu_B}{T}\right)^2 + \mathcal{O}\left(\left(\frac{\mu_B}{T} \right)^4\right) \rm,
\end{equation}
where $c_0$ is the pressure at vanishing chemical potentials and $c_2$ is a mixture of second-order susceptibilities
\begin{equation}\label{eq:c2}
    c_2 = \frac{1}{2}\chi_{BB} + \frac{1}{3} \chi_{BS} + \frac{1}{18} \chi_{SS} \rm .
\end{equation}
We note that the coefficient $c_2$ is evaluated at vanishing chemical potentials. At a sufficiently small net-baryon density, the strangeness neutrality condition can be approximated by $\mu_S = 0$. Thus, one can expect that $\chi_{BB} \approx 2c_2$, as is found to be the case in the HRG model. This allows us to compile the LQCD data on $\chi_{BB}$~\cite{Ding:2023bft} and $c_2$~\cite{Astrakhantsev:2024mat} in a single figure. They are shown in Fig.~\ref{fig:c2_comp} for $T=0.145$ and $0.155~$GeV. We find that the HRG model systematically underestimates the LQCD data for $eB \leq 0.16~\rm GeV^2$ at both temperatures. We note that this is in contrast to the HRG model with an augmented list of resonances which includes states predicted by relativistic quark models. Such a HRG model reproduces $\chi_{BB}$ at vanishing magnetic field~\cite{Bollweg:2021vqf}, but, as the magnetic field increases, both the HRG model with established states only, and the model with all the predicted states, start to overestimate the LQCD results. The calculated susceptibility increases rapidly and drastically overshoots the LQCD data at $eB = 0.3~\rm GeV^2$. The contribution of baryons with $s\geq 3/2$ ranges from $75\%$ at vanishing $eB$ to $95\%$ at $eB=0.3~\rm GeV^2$. This result exposes the weakness of the current treatment of the higher-spin resonances. 

Ultimately, to fully evaluate the contributions of higher-spin states to thermodynamic observables their internal structure should be taken into account. 
This, however, requires a dynamical treatment of resonances, e.g. the S-matrix formulation of statistical mechanics~\cite{Lo:2017ldt, Lo:2017lym, Cleymans:2020fsc}, suitably extended to the presence of $eB$, and is currently under development. 
As a temporary attempt, we consider the detailed balance approach to neutral particles (see the next Section).

Based on our results and arguments presented above, we conclude that the current uncertain treatment of high-spin particles in a finite magnetic field prevents a direct comparison of the HRG model with both the experimental data from HICs and results obtained in LQCD simulations.

\begin{figure}[t!]
    \centering
    \includegraphics[width=\columnwidth]{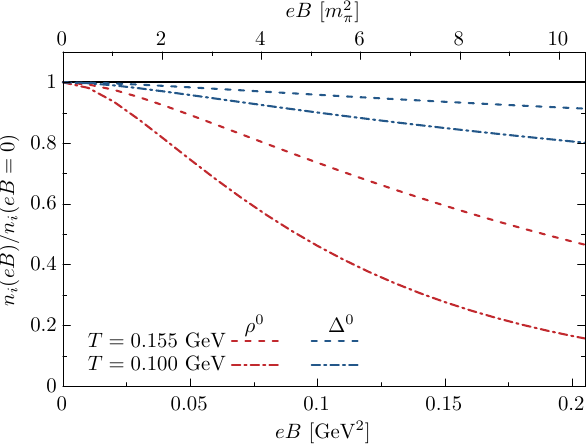}
    \caption{$\rho^0$ and $\Delta^0$ densities when the detailed balance is required, normalized to densities at $eB=0$, at $T = 0.1$ and 0.155 GeV as functions of the magnetic field.}
    \label{fig:density_detailed}
\end{figure}

\section{Detailed Balance}
\label{sec:det_balance}

So far, we have assumed neutral particles to have zero magnetic moments. Consequently they were not affected by the magnetic field at all. However, in chemically equilibrated HRG detailed balance prevails, e.g.,~the decay rate of rho mesons is equal to the scattering rate of pions forming $\rho$ resonances. The scattering rate depends on the densities and, as seen in Fig.~\ref{fig:density_eB}, pion density decreases with the magnetic field. Thus, the $\rho^0$ density may depend on the magnetic field after all.

However, systematic calculation of the decay rates of resonances is challenging. For example, the pair production in the decay process of a neutral scalar boson into two charged scalar bosons is suppressed by the presence of external magnetic field~\cite{Piccinelli:2017yvl}, a neutral boson emission rate by a fermion is enhanced by the magnetic field~\cite{Jaber-Urquiza:2023sal}. Nevertheless, it is known that a weak magnetic field causes a negligibly small change in the width of $\rho^0$~\cite{Bandyopadhyay:2016cpf, Ghosh:2019fet}. Thus, we make an approximation that in the presence of a magnetic field, the $\rho^0$ decay rate changes only if the $\rho^0$ density changes, and the change in the rate is proportional to the change in density.

Unfortunately, we cannot evaluate the scattering rate of pions using a conventional kinetic theory calculation, since particles in magnetic field do not have well-defined momenta. To gain insight into how the detailed balance requirement could affect the densities of neutral resonances, we make the bold assumption that the magnetic field affects the pion scattering rate (and thus $\rho^0$ production rate) only by changing their densities. In other words, at fixed temperature,
\begin{equation}\label{eq:rho_production}
    \Gamma(\pi^+ \pi^- \rightarrow \rho^0) = \frac{n_{\pi^+}(B)\,n_{\pi^-}(B)}{n^{eq}_{\pi^+}\,n^{eq}_{\pi^-}}\Gamma^{eq}(\pi^+\pi^-\rightarrow \rho^0),
\end{equation}
where $\Gamma(\pi^+ \pi^- \rightarrow \rho^0)$ is the production rate of $\rho^0$ in the presence of a magnetic field, $n_{\pi^\pm}(B)$ are the pion densities in magnetic field, $n^{eq}_{\pi^\pm}$ are the corresponding pion densities when $eB =0$, and $\Gamma^{eq}(\pi^+ \pi^- \rightarrow \rho^0)$ the $\rho^0$ production rate in equilibrated HRG when there is no magnetic field.

\begin{figure}[t!]
    \centering
    \includegraphics[width=\columnwidth]{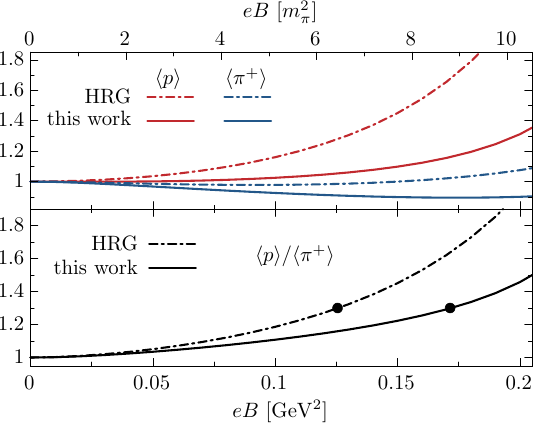}
    \caption{Proton and pion yields (top panel) and their ratio (bottom panel) as functions of magnetic field $eB$ calculated using the conventional HRG in magnetic field approach (HRG) and using the requirement of detailed balance discussed in Sect.~\ref{sec:det_balance} (this work). The results are for $T = 0.155~\rm GeV$ and are normalized to their values at vanishing magnetic field.}
    \label{fig:densityDecay_detailed}
\end{figure}

We can evaluate an effective pion chemical potential such that
\begin{equation}\label{eq:mu_pi}
    n_\pi(T,\mu_\pi,B=0) = n_\pi(T,B).
\end{equation}
With the above-mentioned considerations, detailed balance requires that $\rho^0$ obtains a chemical potential, which is a sum of pion chemical potentials: $\mu_{\rho^0} = \mu_{\pi^+} + \mu_{\pi^-} = 2\mu_{\pi^+}$, since both $\pi^+$ and $\pi^-$ are affected the same way by the magnetic field, and thus their effective chemical potentials are equal. We can repeat similar arguments for $\Delta^0$ resonance, and obtain a requirement that it also develops an effective chemical potential in magnetic field: $\mu_{\Delta^0} = (\mu_{p} + \mu_{\pi^+})/3$.

We have evaluated the $\rho^0$ and $\Delta^0$ densities corresponding to these chemical potentials, and show them scaled with their equilibrium densities in Fig.~\ref{fig:density_detailed}. Unlike the expectation that neutral particles should not be affected by the magnetic field, both $\rho^0$ and $\Delta^0$ densities decrease with increasing strength of magnetic field, the effect being the stronger the colder the system. We note here that our assumption about the production rate is extremely naive, but the $\rho^0$ and $\Delta^0$ densities staying independent of the magnetic field would require very careful fine-tuning of the pion-pion and pion-nucleon interactions.

Detailed balance must prevail not only between neutral resonances and their daughters, but also between charged resonances and their daughters. To gauge the overall uncertainty in the chemical composition of HRG in magnetic field, we assume that the densities of $\Delta$ and $\rho$ are not given by Eqs.~\ref{eq:energy_B}--\ref{eq:density}, but all charged states of $\Delta$ and $\rho$ gain effective chemical potentials similar to $\Delta^0$ and $\rho^0$. We keep all the other densities unchanged and evaluate the pion and proton yields after decays. The resulting yields and their ratios are shown in Fig.~\ref{fig:densityDecay_detailed}. As expected, the reduction in $\rho$ and $\Delta$ densities leads to reduced yields, and since the reduction in $\langle p \rangle$ is greater than in $\langle \pi^+\rangle$, the proton-to-pion ratio becomes smaller.

In the literature, particle ratios have been used to constrain the magnetic field in heavy-ion collisions (e.g.,~\cite{Xu:2020sui, dePaoli:2012cz}). Here we apply this idea to the proton-to-pion ratio in particular. We assume that the field is negligible in most central collisions, largest in most peripheral, and that the centrality dependence of the $p/\pi$ ratio is entirely due to the change in the magnetic field. The ALICE collaboration has measured the $p/\pi$ ratio in Pb+Pb collisions at $\sqrt{s_{NN}} = 5.02~\rm TeV$~\cite{ALICE:2019hno}, and reported that the ratio increases by a factor $\sim 1.3$ from central to peripheral collisions. In the lower panel of Fig.~\ref{fig:densityDecay_detailed}, we have included dots on the $p/\pi$ curves indicating where the ratio has increased by 30\% compared to its value at vanishing magnetic field. We find that this occurs at $eB = 0.13~{\rm GeV^2}\simeq 6.6~m_\pi^2$ and $eB = 0.20~{\rm GeV^2}\simeq 10.1~m_\pi^2$. Thus, our assumptions about the chemical composition of HRG lead to a $\sim 30$\% uncertainty in the allowed strength of the magnetic field.

\section{Conclusions}
\label{sec:conclusions}

In this work, we examined the effect of finite magnetic field on the thermal properties of hadronic medium. We modeled the hadronic phase of QCD with the hadron resonance gas (HRG) model. For spin-$0$ particles, the densities decrease with magnetic field, while the densities of spin-$1/2$ show the opposite behavior which is due to a different structure of the lowest Landau level. The densities of higher-spin resonances strongly increase with magnetic field which is caused by a singular behavior of the dispersion relation~(cf. Eq.~\eqref{eq:energy_B}) applied to higher-spin particles. Since these densities contribute to final production yields of proton and pions after including the resonance decays, the latter also diverge at sufficiently large fields. This may be problematic for a reliable estimation of magnetic field from non-central HICs based on observed particle yields. As an example, we compared the magnetic field predicted using the HRG and the detailed balance conditions and found a substantial difference between both predictions. Our results suggest that the interpretation of the magnetic field dependence of particle yields in the HRG model, although feasible, should be treated with great care.

We have also studied the second-order fluctuations of conserved charges in conjunction with the recent LQCD data. We find that for our choice of the resonance content of HRG, the LQCD data is underestimated by model results at small fields, while the singular behavior at larger $eB$ leads to overestimation of the data. Although including hypothetical states improves the description of LQCD data at small $eB$~\cite{Ding:2023bft}, it makes the breakdown in strong fields even worse. 

Therefore, for a reliable comparison between LQCD results and experimental data, it would be crucial to treat the dynamical structure of resonances, especially those with higher spins and neutral states. As a first step, we explored the effect of the principle of detailed balance. Another possibility is to generalize the S-matrix approach to finite magnetic field which would allow us to take into account the dynamics of hadrons via the scattering data. Ultimately, non-perturbative approaches capable of describing bound states should be employed to study the dynamics of hadrons in an external magnetic field. Many of these interesting points will be pursued in future research.

{\it{Note added}}. An article discussing similar issues appeared simultaneously with the present work~\cite{Vovchenko:2024wbg}.

\section*{Acknowledgments}
This work is supported by the program Excellence Initiative–Research University of the University of Wroc\l{}aw of the Ministry of Education and Science (P.H, M.M. and B.K.). M.S. acknowledges the financial support of the Polish National Science Center (NCN) under the Preludium grant 2020/37/N/ST2/00367. This work is supported partly by the Polish National Science Centre (NCN) under OPUS Grant No. 2022/45/B/ST2/01527 (C.S., K.R. and P.M.L.). K.R. also acknowledges the support of the Polish Ministry of Science and Higher Education. The work of C.S. was supported in part by the World Premier International Research Center Initiative
(WPI) under MEXT, Japan.

\bibliography{biblio}

\end{document}